\documentclass[pre,superscriptaddress,twocolumn]{revtex4}

\usepackage{graphicx}
\usepackage{dcolumn}
\usepackage{array}
\usepackage{tabularx}
\usepackage{color}

\usepackage{floatrow}

\usepackage{eucal}
\usepackage[dvips]{epsfig}
\usepackage{amssymb}
\usepackage{amsfonts}

\usepackage{amsmath,amsthm}

\newcommand{\be}{\begin{eqnarray}}
\newcommand{\ee}{\end{eqnarray}}
\newcommand{\bse}{\begin{subequations}}
\newcommand{\ese}{\end{subequations}}

\newcommand{\bnum}{\begin{enumerate}}
\newcommand{\enum}{\end{enumerate}}

\newcommand{\bit}{\begin{itemize}}
\newcommand{\eit}{\end{itemize}}

\newcommand{\bc}{\begin{cases}}
\newcommand{\ec}{\end{cases}}

\newcommand{\bpm}{\begin{pmatrix}}
\newcommand{\epm}{\end{pmatrix}}

\newcommand{\bvm}{\begin{vmatrix}}
\newcommand{\evm}{\end{vmatrix}}

\newcommand{\bs}{\boldsymbol}

\newcommand{\mcal}{\mathcal}

\newcommand{\gl}{\lambda}
\newcommand{\gk}{\kappa}

\newcommand{\gs}{\sigma}

\newcommand{\Gc}{\Gamma}

\newcommand{\Gl}{\Lambda}

\newcommand{\p}{\partial}
\newcommand{\f}{\frac}

\newcommand{\lan}{\langle}
\newcommand{\ran}{\rangle}

\newcommand{\tn}{\textnormal}

\begin{document}

\title{Spontaneous mirror-symmetry breaking induces inverse energy cascade\\
 in 3D active fluids}

\author{Jonasz S\l{}omka}
\affiliation{Department of Mathematics, Massachusetts Institute of Technology, 77 Massachusetts Avenue, Cambridge, MA 02139-4307, USA}

\author{J{\"o}rn Dunkel}
\affiliation{Department of Mathematics, Massachusetts Institute of Technology, 77 Massachusetts Avenue, Cambridge, MA 02139-4307, USA}

\date{\today}
\begin{abstract}
Classical turbulence theory assumes that energy transport in a 3D turbulent flow proceeds through a Richardson cascade whereby larger vortices successively decay into smaller ones. By contrast, an additional inverse cascade characterized by vortex growth exists in 2D fluids and gases, with profound implications for meteorological flows and fluid mixing. The possibility of a helicity-driven inverse cascade in 3D fluids had been rejected in the 1970s based on equilibrium-thermodynamic arguments. Recently, however, it was proposed that certain symmetry-breaking processes could potentially trigger a 3D inverse cascade, but no physical system exhibiting this phenomenon has been identified to date. Here, we present analytical and numerical evidence for the existence of an inverse energy cascade in an experimentally validated 3D active fluid model, describing microbial suspension flows that spontaneously break mirror symmetry. We show analytically that self-organized scale selection, a generic feature of many biological and engineered nonequilibrium fluids, can generate parity-violating Beltrami flows. Our simulations further demonstrate how active scale selection controls mirror-symmetry breaking and the emergence of a 3D inverse cascade.
\end{abstract}


\maketitle





Turbulence, the chaotic motion of liquids and gases, remains one of the most widely studied phenomena in classical physics~\cite{2004Frisch,McComb_Book}. Turbulent flows determine energy transfer and material mixing over a vast range of scales, from the interstellar medium~\cite{1984Higdon,federrath2010comparing} and solar winds~\cite{2013Bruno_LivRev} to the Earth's atmosphere~\cite{nastrom1984kinetic,1999Lindborg}, ocean currents~\cite{2005Thorpe_OceanBook}, and our morning cup of coffee. Of particular recent interest is the interplay of turbulence and active biological matter~\cite{Enriquez01082015}, owing to its relevance for carbon fixation and nutrient transport in marine ecosystems~\cite{taylor2012trade}. Although much has been learned about the statistical and spectral properties of turbulent flows both experimentally~\cite{1997Noullez_JFM,1999LewisSwinney_PRE,2006Bodenschatz} and theoretically~\cite{kolmogorov1941local,1941Kolmogorov_2,1941Kolmogorov_3,kraichnan1967inertial,1980Kraichnan,waleffe1992,2002Kellay,PhysRevLett.116.124502} over the last 75 years, several fundamental physical and mathematical~\cite{Fefferman_Clay}  questions still await their answer. One of the most important among them, with profound implications for the limits of hydrodynamic mixing, concerns whether 3D turbulent flows can develop an inverse cascade that transports energy from smaller to larger scales~\cite{kraichnan1973helical,waleffe1992,biferale2012inverse}.

\par
Kolmogorov's 1941 theory of turbulence~\cite{kolmogorov1941local} assumes that turbulent energy transport in 3D proceeds primarily from larger to smaller scales through the decay of vortices. This forward (Richardson) cascade is a consequence of the fact that the 3D inviscid Euler equations conserve energy~\cite{2004Frisch}. In 1967, Kraichnan~ \cite{kraichnan1967inertial} realized that the presence of a second conserved quantity, enstrophy, in 2D turbulent flows implies the existence of two dual cascades~\cite{2012Bofetta}: a vorticity-induced cascade to smaller scales and an inverse energy cascade to larger scales~\cite{2000Danilov,2002Kellay}. Two years later, Moffatt~\cite{moffatt1969degree} discovered a new invariant of the 3D Euler equations, which he termed helicity. Could helicity conservation generate an inverse turbulent cascade in 3D? Building on thermodynamic considerations, Kraichnan~\cite{kraichnan1973helical}  argued in 1973 that this should not be possible, but he also conceded that turbulent flows do not necessarily follow equilibrium statistics. Since then, insightful theoretical studies~\cite{waleffe1992,biferale2012inverse} have elucidated other important conditions for the emergence of helicity-driven inverse cascades in 3D fluids, in particular identifying mirror-symmetry breaking as a key mechanism~\cite{biferale2012inverse}. However, no natural or artificially engineered fluid system exhibiting this phenomenon has been identified to date.

\begin{figure*}[t!]
\centering
\includegraphics[width=0.9\textwidth]{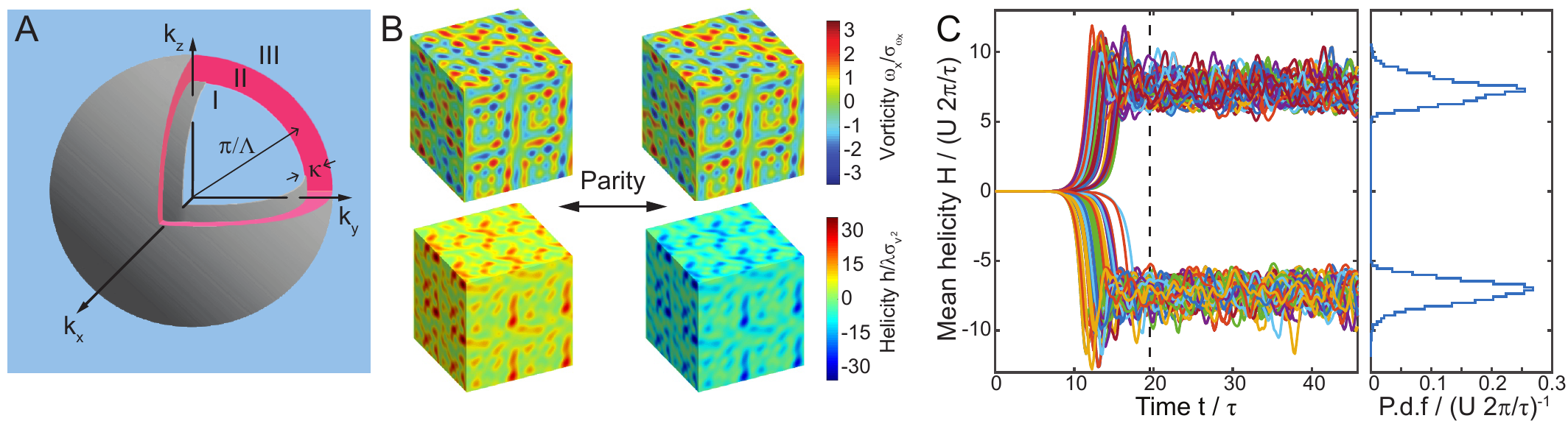}
 \caption{Exact Beltrami-flow solutions and spontaneous mirror symmetry breaking in 3D simulations.
(A)~Linear stability analysis of Eqs.~\eqref{e:eom} distinguishes three different regions in Fourier space: Domains I and III are dissipative, whereas II represents active modes. The radius of the active shell II corresponds approximately to the inverse of characteristic pattern formation scale~$\Lambda$. The bandwidth $\kappa$ measures the ability of the active fluid component to concentrate power input in Fourier space.
(B)~Two examples of exact stationary bulk solutions of Eqs.~\eqref{e:eom} realizing Beltrami vector fields of opposite helicity, obtained from Eq.~\eqref{e:exact_Beltrami} by combining modes of the same helicity located on one of the marginally stable grey surfaces in (A).
(C)~Simulations with random initial condition spontaneously select one of two helicity branches. The histogram represents an average over $150$ runs with random initial conditions, sampled over the statistically stationary state starting at time $t=20\tau$ (dashed line).  Simulation parameters: $\Lambda=75\,\mu$m, $U=72\,\mu$m/s, $\kappa_\tn{I}=0.9/\Lambda$, $L=8\Lambda$ (see also Fig.~\ref{fig02} and SI for larger simulations).
}
\label{fig01} 
\end{figure*}

Here, we predict that fluid flows in active nonequilibrium liquids, such as bacterial suspensions, can spontaneously break mirror symmetry, resulting in a 3D inverse cascade. Broken mirror symmetry plays an important role in nature, exemplified by the parity-violating weak interactions~\cite{1957PhRv..105.1413W} in the standard model of particle physics, by the helical structure of DNA~\cite{1953Natur.171..737W} or, at the macroscale, by chiral seed pods~\cite{Armon1726}. Another, fluid-based realization~\cite{Woodhouse2012_PRL} of a spontaneously broken chiral symmetry was recently observed in confined bacterial suspensions~\cite{Wioland2013_PRL,2016Wioland_NPhys}, which form stable vortices of well-defined circulation when the container dimensions match the correlation scale~$\sim 70\,\mu$m of the collective cell motion in bulk~\cite{2012Sokolov,Dunkel2013_PRL}. Motivated by these observations, we investigate a generalized Navier--Stokes model~\cite{2015SlomkaDunkel,2016SlomkaDunkel} for pattern-forming nonequilibrium fluids that are driven by an active component, which could be swimming bacteria~\cite{2012Sokolov,Dunkel2013_PRL} or ATP-driven microtubules~\cite{2012Sanchez_Nature,2015Giomi} or artificial microswimmers~\cite{Bricard:2013aa,2008Walther_SM,2011Shashi}. The theory uses only generic assumptions about the symmetries and long-wavelength structure of active stress tensors and captures the experimentally observed bulk vortex dynamics in 3D bacterial suspensions~\cite{2012Sokolov,Dunkel2013_PRL} and in flows driven by isotropic active microtubule networks~\cite{2012Sanchez_Nature} (Fig.~S1).

\par
To demonstrate the existence of a helicity-driven inverse cascade in 3D active bulk fluids, we first verify analytically the existence of exact parity-violating Beltrami-flow~\cite{arnold1965topologie,dombre1986chaotic,etnyre2000contact} solutions. We then confirm numerically that active bulk flows starting from random initial conditions approach attractors that spontaneously break mirror symmetry and are statistically close to Beltrami-vector fields. Finally, we demonstrate that the broken mirror symmetry leads to an inverse cascade with triad interactions as predicted by Waleffe~\cite{waleffe1992} about 25 years ago.

\section*{RESULTS}
\textbf{Theory.} 
We consider pattern-forming nonequilibrium fluids consisting of a passive solvent component, such as water, and a stress-generating active component, which could be bacteria~\cite{2012Sokolov}, ATP-driven microtubules~\cite{2012Sanchez_Nature}, or Janus particles~\cite{PhysRevLett.105.268302,2013Buttinoni_PRL}. In contrast to earlier studies, which analyzed the velocity field of the active matter component~\cite{2012Wensink,Dunkel2013_PRL,Bratanov08122015}, we focus here on the incompressible solvent velocity field $\bs v(t,\bs x)$ described by
\bse
\label{e:eom}
\be
\label{e:eoma}
\nabla\cdot\bs v&=&0, \\
\label{e:eomb}
\p_t \bs v+\bs v \cdot \nabla \bs v&=&-\nabla p+\nabla \cdot{\bs \sigma},
\ee
where   $p(t,\bs x)$ is the local pressure.  The effective stress tensor $\bs\gs(t,\bs x)$ comprises passive contributions from the intrinsic solvent fluid viscosity and active contributions representing the stresses exerted by the microswimmers on the fluid~\cite{2002Ra,2008SaintillanShelley,2013Marchetti_Review,2013Ravnik_PRL}. Experiments~\cite{2004DoEtAl,2007Cisneros,2012Sanchez_Nature,2012Sokolov,Dunkel2013_PRL,Wioland2013_PRL} show that  active stresses typically lead 
to vortex scale selection in the ambient solvent fluid. This mesoscale pattern-formation stands in stark contrast to the scale-free vortex structures in externally driven classical fluid turbulence and can be described  phenomenologically through the stress tensor~\cite{2015SlomkaDunkel,2016SlomkaDunkel}
\be
\label{e:stress}
\bs\gs=(\Gamma_0 -\Gamma_2 \nabla^2+\Gamma_4 \nabla^4)[\nabla \bs v+ (\nabla \bs v)^\top],
\ee
\ese
where the higher-order derivatives $\nabla^{2n}\equiv (\nabla^2)^n$, \mbox{$n\ge2$} account for non-Newtonian effects~\cite{2014BipolarBook} (SI \textit{Model Justification}).  Such higher-order stresses arise naturally from diagrammatic expansions~\cite{2011Ma}. Similar 1D and 2D models have been studied in the context of soft-mode turbulence and seismic waves~\cite{1993BeNi_PhysD,1996Tribelsky_PRL,PhysRevE.77.035202}. 
\par
The parameters $(\Gamma_0,\Gamma_2,\Gamma_4)$ encode microscopic interactions, thermal and athermal fluctuations, and other nonequilibrium processes. For $\Gc_2=\Gc_4=0$, Eqs.~\eqref{e:eom} reduce to the standard Navier-Stokes equations  with kinematic viscosity $\Gamma_0>0$.  For $\Gc_0>0,\Gc_4>0$ and $\Gc_2<0$, Eq.~\eqref{e:stress} defines the simplest ansatz for an active stress tensor that is isotropic, selects flow patterns of a characteristic scale (Fig.~\ref{fig01}), and yields a stable theory at small and large wavenumbers~\cite{2015SlomkaDunkel,2016SlomkaDunkel}. The active-to-passive phase transition corresponds to a sign change from $\Gamma_2<0$ to $\Gamma_2> 0$, which can be realized experimentally through ATP or nutrient depletion. The non-negativity of  $\Gamma_0$ and $\Gamma_4$ follows from stability considerations. $\Gamma_0$ describes the damping of long-wavelength  perturbations  on scales much larger than the typical correlation length of the coherent flow structures, whereas $\Gamma_{2}$ and $\Gamma_4$ account for the growth and damping of modes at intermediate and small scales~(Fig.~\ref{fig01}A).  The resulting nonequilibrium flow structures can be characterized in terms of the typical vortex size $\Lambda=\pi\sqrt{2\Gamma_4/(-\Gamma_2)}$, growth timescale~\cite{2016SlomkaDunkel} 
$$
\tau=\left[\f{\Gamma_2}{2\Gamma_4}\left(\Gamma_0-\f{\Gamma_2^2}{4\Gamma_4}\right)\right]^{-1},
$$
circulation speed $U=2\pi\Lambda/\tau$ and spectral bandwidth (Fig.~\ref{fig01}A)
$$
\kappa=\biggl(\f{-\Gamma_2}{\Gamma_4}-2\sqrt{\f{\Gamma_0}{\Gamma_4}}\biggr)^{1/2}.
$$ 
Specifically, we find $\Lambda=41\,\mu$m, $U=57\,\mu$m/s and $\kappa=73$ mm$^{-1}$  for flows measured in \textit{Bacillus subtilis} suspensions~\cite{2012Sokolov,Dunkel2013_PRL}  and  $\Lambda=130\,\mu$m, $U=6.5\,\mu$m/s and $\kappa=21$ mm$^{-1}$ for ATP-driven microtubule-network suspensions~\cite{2012Sanchez_Nature}  (SI \textit{Comparison with Experiments}). We emphasize, however, that truncated polynomial stress-tensors of the form~\eqref{e:stress} can provide useful long-wavelength approximations for a broad class  of pattern-forming liquids,  including magnetically~\cite{2008Ouellette}, electrically~\cite{C5SM02316E}, thermally~\cite{PhysRevLett.105.268302,2014Bregulla_ACSNano,2015Fedosov_SoftMatter} or chemically~\cite{2013Loewen_PRL,2013Buttinoni_PRL} driven flows.

\textbf{Exact Beltrami-flow solutions and broken mirror symmetry.}
The higher-order Navier-Stokes equations defined by Eqs.~\eqref{e:eom}  are invariant under the parity transformation $\bs x \to -\bs x$.
Their solutions however, can spontaneously break this mirror symmetry. To demonstrate this explicitly, we construct a family of exact nontrivial stationary solutions in free space by  decomposing the Fourier series $\bs v (t,\bs k)$ of the divergence-free velocity field $\bs v(t,\bs x)$ into helical modes~\cite{waleffe1992,biferale2012inverse}
\be
\label{e:vel2hmodes}
\bs v (t,\bs k)=u^+(t,\bs k)\,\bs h^+(\bs k)+u^-(t,\bs k)\,\bs h^-(\bs k),
\ee
where $\bs h^\pm$ are the eigenvectors of the curl operator, \mbox{$i\bs k \wedge \bs h^\pm=\pm k \bs h^\pm$} with $k=|\bs k|$. 
Projecting Eq.~(\ref{e:eomb}) onto helicity eigenstates \cite{waleffe1992} yields the evolution equation for the mode amplitudes $u^\pm$
\be
\label{e:eomh}
[\p_t+\xi(k)]u^\pm (t,\bs k)=\sum_{(\bs p,\bs q)\;:\;\bs k+\bs p +\bs q=0}f^\pm (t; \bs k, \bs p, \bs q),
\ee
where $\xi(k)=\Gamma_0 k^2+\Gamma_2 k^4+\Gamma_4 k^6$ is the active stress contribution, and the nonlinear advection is represented by all triadic interactions~\cite{waleffe1992,biferale2012inverse} 
\be
\label{e:advectionFourier}
 f^{s_k}  (t; \bs k, \bs p, \bs q) &=&-\f{1}{4}\sum_{s_p, s_q}(s_p p-s_q q)
 \times\\
 \notag &&\qquad\qquad
 \left[(\overline{\bs h}^{s_p}\wedge \overline{\bs{h}}^{s_q})\cdot \overline{\bs{ h}}^{s_k}\right]
 \overline{u}^{s_p}\overline{u}^{s_q}
 \ee
between helical $\bs k$-modes and $\bs p,\, \bs q$-modes, where $s_k,\, s_p, s_q\in\{\pm\} $ are the corresponding helicity indices (overbars denote complex conjugates of ${\bs h}^{s_p}={\bs h}^{s_p}(\bs p)$, etc. \cite{waleffe1992}). There are two degrees of freedom per wavevector, and hence eight types of interactions for every triple $(\bs k,\bs p, \bs q)$. As evident from Eq.~\eqref{e:advectionFourier}, arbitrary superpositions of modes with identical wavenumber $p=q=k_*$ and same helicity index annihilate the advection term, because $s_p p-s_q q=0$ in this case. Therefore, by choosing  $k_*$ to be a root of the polynomial $\xi(k)$, corresponding to the grey surfaces in Fig.~\ref{fig01}A, we obtain exact stationary solutions
\be
\label{e:exact_Beltrami}
\bs v^\pm (\bs x)=\sum_{\bs k, k=k_*}u^\pm(\bs k)\bs h^\pm(\bs k)e^{i\bs k\cdot\bs x},
\ee
where $u^\pm (-\bs k)=\overline{ u}^\pm (\bs k)$ ensures real-valued flow fields. In particular, these solutions~\eqref{e:exact_Beltrami} correspond to Beltrami flows~\cite{arnold1965topologie,dombre1986chaotic,etnyre2000contact}, obeying $\nabla\wedge \bs v^\pm=\pm k_* \bs v^\pm$. Applying the parity operator to any right handed solution $\bs v^+(\bs x)$ generates the corresponding left handed solution $\bs v^-(\bs x)$ and \textit{vice versa} (Fig.~\ref{fig01}B). 
\par
Although the exact solutions $\bs v^\pm (\bs x)$ describe stationary Beltrami fields~\cite{arnold1965topologie,dombre1986chaotic,etnyre2000contact} of fixed total helicity \mbox{$\mcal{H}^\pm=\int d^3x\, \bs v^\pm\cdot \bs\omega^\pm$}, where $\bs \omega=\nabla\wedge \bs v$ is the vorticity, it is not yet clear whether parity violation is a generic feature of arbitrary time-dependent solutions of Eqs.~\eqref{e:eom}. However, as we will demonstrate next, simulations with random initial conditions do indeed converge to  statistically stationary flow states that spontaneously break mirror symmetry and are close to Beltrami flows.

\begin{figure*}[t!]
\centering
\includegraphics[width=0.9\textwidth]{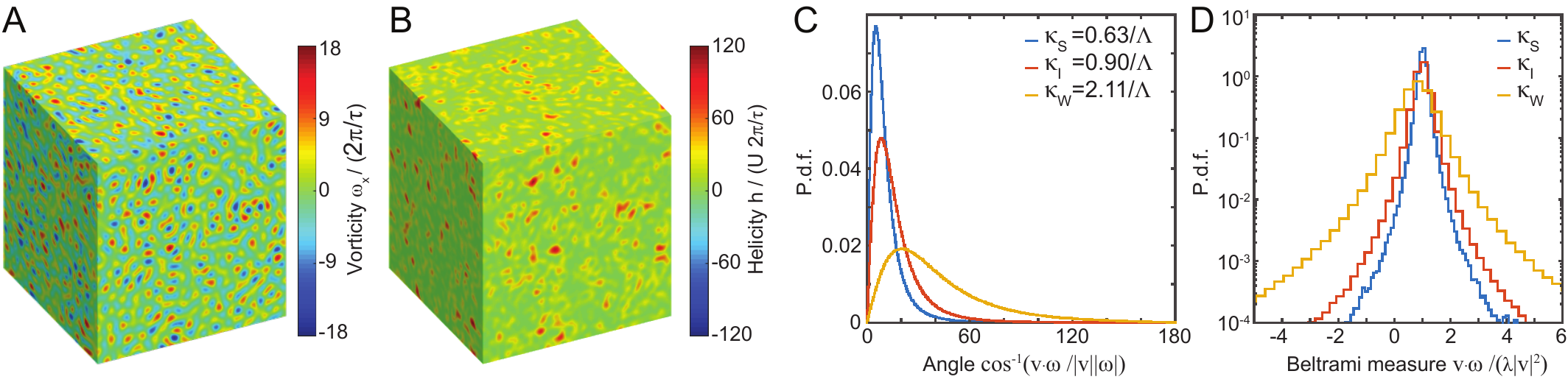}
 \caption{
Active fluids spontaneously break mirror symmetry by realizing Beltrami-type flows.
(A)~Snapshot of a representative vorticity component field  $\omega_x$ (Movie 1)  for an active fluid with small bandwidth $\kappa_\tn{S}=0.63/\Lambda$, as defined in Fig.~\ref{fig01}A.
(B)~The corresponding helicity field signals parity-symmetry breaking leading to a positive-helicity flow in this example.
(C)~Histograms of the angles between velocity $\bs v$ and vorticity $\bs \omega$ quantify the alignment between the two fields for different active bandwidths $\kappa_\tn{S}<\kappa_\tn{I}<\kappa_\tn{W}$: The smaller the bandwidth, the stronger the alignment between $\bs v$ and~$\bs \omega$.
(D)~Numerically estimated distributions of the Beltrami measure, $\beta=\bs v \cdot \bs \omega / (\lambda|\bs v|^2)$, shown on a log-scale. An ideal Beltrami flow with $\bs \omega=\lambda \bs v$ produces a delta-peak centered at $\beta=1$. Identifying  $\lambda$ with the midpoint of the active shell ($\lambda\approx\pi/\Lambda$), which approximately corresponds to the most unstable wavenumber and the characteristic pattern formation scale, we observe that a smaller active bandwidth leads to a sharper peak and hence more Beltrami-like flows. Data were taken at a single representative time-point long after the characteristic relaxation time. Simulation parameters: $\Lambda=75\,\mu$m, $U=72\,\mu$m/s, $L=32\Lambda$.}
\label{fig02} 
\end{figure*}


\textbf{Spontaneous mirror symmetry breaking in time-dependent solutions.}
We simulate the full nonlinear Eqs.~\eqref{e:eom} on a periodic cubic domain (size $L$) using a spectral algorithm (SI \textit{Numerical Methods}). Simulations are performed for typical bacterial parameters $(\Gamma_0,\Gamma_2,\Gamma_4)$, keeping the vortex scale $\Lambda=75\,\mu$m and circulation speed $U=72\,\mu$m/s fixed~\cite{2012Sokolov,Dunkel2013_PRL} and comparing three different spectral  bandwidths $\kappa_\tn{S}=0.63/\Lambda=8.4\,$mm$^{-1}$, $\kappa_\tn{I}=0.90/\Lambda=12\,$mm$^{-1}$ and $\kappa_\tn{W}=2.11/\Lambda=28.1\,$mm$^{-1}$, corresponding to active fluids with a small (S), intermediate (I) and wide (W) range of energy injection scales. A small bandwidth means that the active stresses inject energy into a narrow shell in Fourier space, whereas a wide bandwidth means energy is pumped into a wide range of Fourier modes (Fig.~\ref{fig01}A). All simulations are initiated with weak incompressible random flow fields. For all three values of $\kappa$, we observe spontaneous mirror-symmetry breaking indicated by the time evolution of the mean helicity $H=(1/L^3)\int d^3x \,h$, where $h=\bs v\cdot \bs\omega$ is the local helicity. During the initial relaxation phase, the flow dynamics is attracted to states of well-defined total helicity and remains in such a statistically stationary configuration for the rest of the simulation.  As an illustration, Fig.~\ref{fig01}C shows results from 150 runs for $\kappa=\kappa_\tn{I}$ and $L=8\Lambda$, with flow settling into a positive (negative) mean helicity state 72 (78) times. This spontaneous mirror-symmetry breaking is robust against variations of the bandwidth and simulation box size, as evident from the local vorticity and helicity fields for $\kappa=\kappa_\tn{S}$ and $L=32\Lambda$ in Fig.~\ref{fig02}A,B.

\textbf{Beltrami-flow attractors.}
Having confirmed spontaneous parity violation for the time-dependent solutions of Eqs.~\eqref{e:eom}, we next characterize the chaotic flow attractors. To this end, we measure and compare the histograms of the angles between the local velocity field $\bs v(t,\bs x)$ and vorticity field $\bs \omega(t,\bs x)$ for the three bandwidths $\kappa_\tn{S}<\kappa_\tn{I}<\kappa_\tn{W}$. Our numerical results reveal that a smaller active bandwidth, corresponding to a more sharply defined scale selection, causes a stronger alignment of the two fields (Fig.~\ref{fig02}C). Recalling that perfect alignment, described by $\bs \omega=\gl \bs v$ with eigenvalue $\lambda$, is the defining feature of Beltrami flows~\cite{arnold1965topologie,dombre1986chaotic,etnyre2000contact}, we introduce the Beltrami measure $\beta=\bs v\cdot\bs \omega/(\lambda |\bs v|^2)$. For ideal Beltrami fields, the distribution of $\beta$ becomes a delta peak centered at $\beta=1$. Identifying $\lambda$ with the midpoint of the active shell ($\lambda\approx \pi/\Lambda$), which approximately corresponds to the most dominant pattern formation scale in Eqs.~\eqref{e:eom}, we  indeed find that the numerically computed flow fields exhibit $\beta$-distributions that are sharply peaked at $\beta=1$ (Fig.~\ref{fig02}D). Keeping  $\Lambda$ and $U$ constant, the sharpness of the peak increases with decreasing active bandwidth~$\kappa$. These results imply that active fluids with well-defined intrinsic scale selection realize flow structures that are statistically close to Beltrami fields, as suggested by the particular analytical solutions derived earlier.

\begin{figure*}[t!]
\centering
\includegraphics[width=\textwidth]{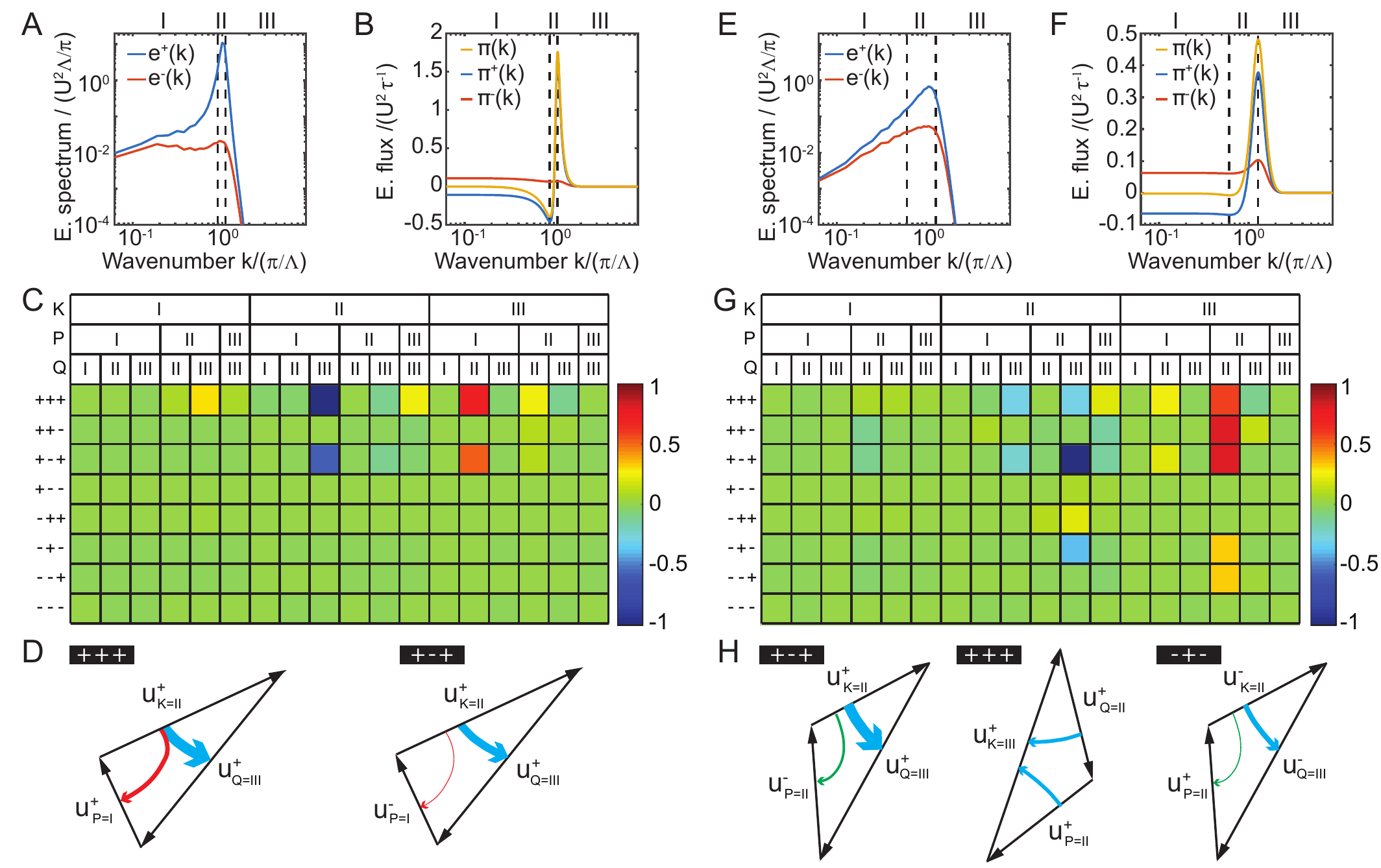}
 \caption{
Scale selection controls mirror symmetry breaking and induces an inverse energy cascade. We demonstrate these effects for active fluids with a (A-D) small active bandwidth $\kappa_\tn{S}$ and (E-H) wide bandwidth $\kappa_\tn{W}$ (Fig.~\ref{fig01}A). The intermediate case $\kappa_\tn{I}$ is presented in Fig.~S3.
(A)~Energy spectra $e^{\pm}(k)$ of the helical velocity-field modes show strong symmetry breaking for small bandwidth parameter $\kappa_\tn{S}$. In this example, the system spontaneously selects positive helicity modes, such that $e^+(k)>e^-(k)$ at all dominant wavenumbers. Dashed vertical lines indicate the boundaries of the energy injection domain II. 
(B)~The resulting energy fluxes $\Pi^{\pm}(k)$ combine into the total flux $\Pi(k)$, which is negative in region I and positive in III, signaling inverse and forward energy cascades, respectively.
(C)~Contributions to the energy flow $\lan \mcal{T}_{KPQ}^{s_K s_P s_Q}\ran$ between the three spectral domains I,II, and III (18 possibilities, columns) from the eight types of triad interactions (rows). In reflection-invariant turbulence, this table remains unchanged under upside-down flipping ($+\leftrightarrow -$). Instead, we observe a strong asymmetry, with two cumulative triads (D) dominating the energy transfer. Red and blue arrows represent transfer towards large and small scales, respectively, and thickness represents magnitude of energy flow. Green arrows represent transfer within the same spectral domain. The direction of the energy flow is in agreement with the instability assumption of Waleffe~\cite{waleffe1992}. In this case, 18.2\% of the injected energy is transferred from region II to region I and 81.8\% is transferred from II to III.
(E-H)~Same plots for an active fluid with wide active bandwidth $\kappa_\tn{W}$. 
(E and F)~Energy spectra show weaker parity breaking (E) and suppression of the inverse energy cascade (F). 
(G)~The energy flow table partially recovers the upside-down ($+\leftrightarrow -$) symmetry. 
(H)~The most active triads now favor the forward cascade, so that  only 1.1\% of the injected energy flows into region I, while 98.9\% are transferred into region III.
Data represent averages over single runs~(Fig.~S2). Simulation parameters are identical to those in Fig.~\ref{fig02}.
}
\label{fig03} 
\end{figure*}

\section*{DISCUSSION}

\textbf{Spontaneous parity breaking \textbf{vs.} surgical mode removal.}
Important previous studies identified bifurcation mechanisms~\cite{1984MalomedTribelsky,coullet1989parity,PhysRevLett.74.4839} leading to parity-violation in 1D and 2D~\cite{2003Fujisaka} continuum models of pattern-forming nonequilibrium systems~\cite{PhysRevA.43.6700,1997Tribelsky_Review}.  The above analytical and numerical results generalize these ideas to 3D fluid flows, by showing that an active scale selection mechanism can induce spontaneous helical mirror-symmetry breaking. Such self-organized parity violation can profoundly affect energy transport and mixing in 3D active fluids, which do not satisfy the premises of Kraichnan's thermodynamic \emph{no-go} argument~\cite{kraichnan1973helical}.  An insightful recent study~\cite{biferale2012inverse}, based on the classical Navier-Stokes equation, found that an \textit{ad hoc} projection of solutions to positive or negative helicity subspaces can result in an inverse cascade but it has remained an open question whether or not such a surgical mode removal can be realized experimentally in passive fluids. By contrast, active fluids spontaneously achieve helical parity-breaking (Fig.~\ref{fig01}C) by approaching Beltrami flow states (Fig.~\ref{fig02}C,D), suggesting the possibility of a self-organized inverse energy cascade even in 3D. Before testing this hypothesis in detail, we recall that the generic minimal model defined by Eqs.~\eqref{e:eom} merely assumes the existence of linear active stresses to account for pattern scale selection as observed in  a wide range of microbial suspensions~\cite{2004DoEtAl,2007SoEtAl,Dunkel2013_PRL,2012Sanchez_Nature}, but does not introduce nonlinearities beyond those already present in the classical Navier-Stokes equations. That is, energy redistribution in the solvent fluid is governed by the advective nonlinearities as in conventional passive liquids.

\textbf{Inverse cascade in 3D active fluids.}
To quantify how pattern scale selection controls parity breaking and energy transport in active fluids, we analyzed large-scale simulations ($L=32 \Lambda$; Fig.~\ref{fig02}A,B) for different values of the activity bandwidth $\kappa$  (Fig.~\ref{fig01}A) while keeping the pattern scale $\Gl$ and the circulation speed $U$ fixed.  The active shell (red domain II in Fig.~\ref{fig01}A) corresponds to the energy injection range in Fourier space and provides a natural separation between large flow scales (blue domain I) and small flow scales (blue domain III). Consequently, the forward cascade corresponds to a net energy flux from domain II to domain III, whereas an inverse cascade transports energy from II to I.  We calculate energy spectra $e(k)=e^+(k)+e^-(k)$  and energy fluxes $\Pi(k)=\Pi^+(k)+\Pi^-(k)$ directly from our simulation data, by decomposing  the velocity field into helical modes as in Eq.~\eqref{e:vel2hmodes}, which yields a natural splitting into cumulative energy and flux contributions  $e^\pm(k)$ and  $\Pi^\pm(k)$ from helical modes $u^\pm (\bs k)$ lying on the wavenumber shell $|\bs k|=k$ (SI \textit{Numerical Methods}). Time-averaged spectra and fluxes are computed for each simulation run after the system has relaxed to a statistically stationary state (Fig.~S2). For a small injection bandwidth $\gk_\tn{S}$, the energy spectra $e^\pm(k)$ reflect the broken mirror-symmetry, with most of the energy being stored in either the  positive-helicity or the negative-helicity modes (Fig.~\ref{fig03}A), depending on the initial conditions. Moreover, in addition to the expected 3D forward transfer, the simulation data for $\gk_\tn{S}$ also show  a significant inverse transfer, signaled by the negative values of the total flux $\Pi(k)$ (yellow curve in Fig.~\ref{fig03}B) in domain I. As evident from 
the blue curves in Fig.~\ref{fig03}A and B, this inverse cascade is facilitated by the helical modes that carry most of the energy. For a large injection bandwidth $\gk_\tn{W}\gg \gk_\tn{S} $, the energy spectra continue to show signatures of helical symmetry breaking (Fig.~\ref{fig03}E), but the energy transported to larger scales becomes negligible relative to the forward cascade, as contributions from opposite-helicity modes approximately cancel in the long-wavelength domain I (Fig.~\ref{fig03}F). Results for the intermediate case $\gk_\tn{M}$  still show a significant inverse transfer (Fig.~S3, S4E), demonstrating how the activity bandwidth -- or, equivalently, the pattern selection range -- controls both parity violation and inverse cascade formation in an active fluid. The upward transfer is non-inertial at intermediate scales, as indicated by the wavenumber dependence of the energy flux (Fig.~\ref{fig03}B). At very large scales $\gg \Gl$, however, the flux approaches an inertial  plateau (SI~\textit{Cascade Characteristics}). In contrast to the energy-mediated 2D inverse cascade in passive fluids, which is characterized by vortex mergers, the helicity-driven 3D inverse cascade in active fluids is linked to the formation of extended vortex chain complexes that move collectively through the fluid (Movie 1 and SI~\textit{Cascade Characteristics}).

\textbf{Triad interactions.} 
Our numerical flux measurements confirm directly the existence of a self-sustained 3D inverse cascade induced by spontaneous parity violation, consistent with earlier projection-based 
arguments for the classical Navier-Stokes equations~\cite{biferale2012inverse}. An inverse energy cascade can exist in 3D active fluids because mirror symmetry breaking favors only a subclass of all possible triad interactions, which describe advective energy transfer in Fourier space between velocity modes $\{\bs v(\bs k),\bs v(\bs p),\bs v(\bs q)\}$ with $\bs k+\bs p+\bs q=0$, cf.~Eq.~\eqref{e:advectionFourier}. To analyze in detail which triads are spontaneously activated in a pattern-forming nonequilibrium fluid, we consider combinations  $K,P,Q\in\{\mathrm{I,II,III}\}$ of the spectral domains in Fig.~\ref{fig01}A and distinguish modes by their helicity index  $s_K, s_P, s_Q\in\{\pm\}$. The helicity-resolved  integrated energy flow into the region $(K,s_K)$ due to interaction with regions $(P,s_P)$ and $(Q,s_Q)$ is given by (SI \textit{Numerical Methods})
\be\label{e:symm_triad}
\mcal{T}_{KPQ}^{s_K s_P s_Q}=\f{1}{2}\big(\tilde{\mcal{T}}_{KPQ}^{s_K s_P s_Q}+\tilde{\mcal{T}}_{KQP}^{s_K s_Q s_P}\big),
\ee
where the unsymmetrized flows are defined by
\be
\label{e:regionhelicalint}
\tilde{\mcal{T}}_{KPQ}^{s_K s_P s_Q}=-\int d^3 x\, \bs v^{s_K}_K\cdot [(\bs v^{s_P}_P\cdot\nabla)\bs v^{s_Q}_Q], 
\ee
with $\bs v^{s_K}_K(t,\bs x)$ denoting the helical Littlewood-Paley velocity components, obtained by projecting on modes of a given helicity index $s_K\in\{\pm\}$ restricted to the Fourier space domain~$K$. Intuitively, entries of $\mcal{T}$ are large when the corresponding triads are dominant.
\par
For active fluids, Fourier space is naturally partitioned into three regions (Fig.~\ref{fig01}A) and there are $2^3=8$ helicity index combinations. The triad tensor $\mcal{T}$ is symmetric in the last two indices, so that that  $\mcal{T}$ has $8\times 18$ independent components encoding the fine-structure of the advective energy transport. Stationary time-averages for $\lan \mcal{T}\ran$, measured directly from our simulations (SI \textit{Numerical Methods}) for small ($\kappa_\tn{S}$) and wide ($\kappa_\tn{W}$) energy injection bandwidths are shown in Fig.~\ref{fig03}C and G. For reflection-symmetric turbulent flows, these two tables would remain unchanged under an upside-down flip~\mbox{({\small$+$} $\leftrightarrow$ {\small$-$})}. By contrast, we find a strong asymmetry for a narrow bandwidth  $\kappa_\tn{S}$  (Fig.~\ref{fig03}C), which persists in weakened form for $\kappa_\tn{W}$ (Fig.~\ref{fig03}G). Specifically, we observe for $\kappa_\tn{S}$ two dominant cumulative triads with energy flowing out of the active spectral range II into the two passive domains I and  III  (Fig.~\ref{fig03}D). These cumulative triads visualize dominant entries of the tables in Figs.~\ref{fig03}C,G and represent  the total contributions from all triadic interactions between modes with given helicity indexes and with \lq legs\rq{} lying in the specified spectral domain. The observed energy transfer directions, with energy flowing out of the intermediate domain II when the small-scale modes carry the same helicity index, are in agreement with a turbulent instability mechanism proposed by Waleffe~\cite{waleffe1992}. Interestingly, however, our numerical results show that both \lq R\rq-interaction channels {\scriptsize$+++$} and {\scriptsize$+-+$} contribute substantially even in the case of strong parity-breaking ($\kappa_\tn{S}$; Fig.~\ref{fig03}D); when one surgically projects the full dynamics onto states with fixed parity, only the {\scriptsize$+++$}  channel remains~\cite{biferale2012inverse}. By contrast, for a wide bandwidth $\kappa_\tn{W}$, the dominating triad interactions (Fig.~\ref{fig03}H) favor the forward cascade. Hence, the inverse energy cascade in a 3D active fluids is possible because only a subset of triadic interactions is active in the presence of strong mirror-symmetry breaking. This phenomenon is controlled by the spectral bandwidth of the scale selection mechanism.

\textbf{Enhanced mixing.}
Eqs.~\eqref{e:eom} describe a 3D isotropic fluid capable of transporting energy from 
smaller to larger scales. Previously,  self-organized inverse cascades were demonstrated only in effectively 2D flows~\cite{1980Kraichnan,2002Kellay,2012Bofetta,nastrom1984kinetic,smith1996crossover,smith1999transfer,smith2002generation,xia2008turbulence,mininni2009scale,celani2010turbulence,xia2011upscale}. The 2D inverse cascade has been intensely studied in meteorology~\cite{nastrom1984kinetic,1999Lindborg}, a prominent example being Jovian atmospheric dynamics~\cite{2005Cassini}, because of its importance for the mixing of thin fluid layers~\cite{PhysRevLett.81.2244,PhysRevLett.98.024501,Bernard:2006aa}.  Analogously, the 3D inverse cascade and the underlying Beltrami-flow structure is expected to enhance mixing and transport in active fluids. Seminal work by Arnold~\cite{arnold1965topologie} showed that steady solutions of the incompressible Euler equations include Beltrami-type ABC flows~\cite{dombre1986chaotic} characterized by chaotic streamlines. Similarly, the Beltrami structure of the active-flow attractors of Eqs.~\eqref{e:eom} implies enhanced local mixing. Combined with the presence of an inverse cascade,  which facilitates  additional large scale mixing  through the excitation of long-wavelength modes, these results suggest that active biological  fluids, such as microbial suspensions~\cite{2004DoEtAl,2007SoEtAl,Dunkel2013_PRL}, can be more efficient at stirring fluids and transporting nutrients than previously thought.

\section*{CONCLUSIONS}

To detect Beltrami flows in biological or engineered active fluids, one has to construct histograms and spectra as shown in Figs.~\ref{fig02}C,D and~\ref{fig03}A,E from experimental fluid velocity and helicity data, which is in principle possible with currently available fluorescence imaging techniques~\cite{2006Bodenschatz,Dunkel2013_PRL}. Moreover, helical tracer particles~\cite{PhysRevFluids.1.054201} can help distinguish left-handed and right-handed flows. The above analysis predicts that Beltrami flow structures, mirror symmetry breaking and the inverse 3D cascade appear more pronounced when the pattern selection is focused in a narrow spectral range. Our simulations further suggest that the relaxation time required for completion of the mirror-symmetry breaking process  depends on the domain size (Fig.~S5). For small systems, the relaxation is  exponentially fast, whereas for large domains relaxation proceeds in two stages, first exponentially and then linearly. In practice, it may therefore be advisable to accelerate relaxation by starting experiments from rotating initial conditions.

\small{
\subsection*{Methods}
Eq.~(\ref{e:eom}) was solved numerically in the vorticity-vector potential form with periodic boundary conditions using a spectral code with 3/2 anti-aliasing (SI~\textit{Numerical Methods}). Tables in Fig.~\ref{fig03} were calculated using the Littlewood--Paley decomposition and collocation.
}


\end{document}